\documentclass[onecolum,secnumarabic,amssymb, nobibnotes,longbibliography,preprint]{revtex4-1}

\usepackage{graphicx,color,amsmath}
\usepackage[normalem]{ulem}

\begin{document}
	
	\title{Motor strategies and adiabatic invariants: The case  of rhythmic motion in parabolic flights }
	
	\author{N. Boulanger$^1$}
	\email[e-mail:]{nicolas.boulanger@umons.ac.be}
	\author{F. Buisseret$^{2,3}$}
	\email[e-mail:]{buisseretf@helha.be}
	\author{V. Dehouck$^{1,4}$}
	\email[e-mail:]{victor.dehouck@alumni.umons.ac.be}
	\author{F. Dierick$^{2,5,6}$}
	\email[e-mail:]{frederic.dierick@gmail.com}
	\author{O. White$^4$}
	\email[e-mail:]{olivier.white@u-bourgogne.fr}
	
	\affiliation{$^1$ Service de Physique de l'Univers, Champs et Gravitation, Universit\'{e} de Mons, UMONS  Research Institute for Complex Systems, Place du Parc 20, 7000 Mons, Belgium }
	\affiliation{$^2$ CeREF, Chaussée de Binche 159, 7000 Mons, Belgium}
	\affiliation{$^3$ Service de Physique Nucl\'{e}aire et Subnucl\'{e}aire, Universit\'{e} de Mons, UMONS Research Institute for Complex Systems, 20 Place du Parc, 7000 Mons, Belgium}
	\affiliation{$^4$ Universit\'{e} de Bourgogne INSERM-U1093 Cognition, Action, and Sensorimotor Plasticity, Campus Universitaire, BP 27877, 21078 Dijon, France}
	\affiliation{$^5$ Facult\'{e} des Sciences de la Motricit\'e, Universit\'e catholique de Louvain, 1 Place Pierre de Coubertin, 1348 Louvain-la-Neuve, Belgium}
		\affiliation{$^6$ Centre National de R\'{e}\'{e}ducation Fonctionnelle et de R\'{e}adaptation -- Rehazenter, Laboratoire d'Analyse du Mouvement et de la Posture (LAMP), Luxembourg, Grand-Duch\'{e} de Luxembourg}

	\date{\today}%

\begin{abstract}
The role of gravity in human motor control is at the same time obvious and difficult to isolate. It can be assessed by performing experiments in variable gravity. We propose that adiabatic invariant theory may be used to reveal nearly-conserved quantities in human voluntary rhythmic motion, an individual being seen as a complex time-dependent dynamical system with bounded motion in phase-space. We study an explicit realization of our proposal: An experiment in which we asked participants to perform $\infty-$ shaped motion of their right arm during a parabolic flight, either at self-selected pace or at a metronome's given pace. Gravity varied between $0$ and $1.8$ $g$ during a parabola. We compute the adiabatic invariants in participant's frontal plane assuming a separable dynamics. It appears that the adiabatic invariant in vertical direction increases linearly with $g$, in agreement with our model. Differences between the free and metronome-driven conditions show that participants' adaptation to variable gravity is maximal without constraint. Furthermore, motion in the participant's transverse plane induces trajectories that may be linked to higher-derivative dynamics. Our results show that adiabatic invariants are relevant quantities to show the changes in motor strategy in time-dependent environments.
\end{abstract}

\maketitle

\section{Introduction}

Gravity obviously plays a role in human motor control, but it is not easy to isolate. On one hand, the perception and internal representation of gravity in the brain results from a widely distributed sensory process, including the vestibular system, vision, and somatosensory information \cite{white20}. On the other hand, the constant immersive nature of the body in the Earth gravitational field calls for holistic experimental approaches in contrast to focused interventions on isolated body parts. In these settings, participants are exposed to variable gravito-inertial fields generated by human centrifuges or parabolic flights. The rationale behind these experimental approaches rest on Einstein's equivalence principle (see \textit{e.g.} \cite{Einstein:1916vd,Wald:106274}): Physics in an accelerated spacecraft (1g) is undistinguishable from physics on the ground. It does, however, not mean that the brain does not attempt to identify the possible different sources that give rise to the same consequences. Beyond the brain's role, it is well-known that variable gravity induces various changes in human physiology, typically at the cardiovascular \cite{Aubert16}, neural \cite{White16} and musculoskeletal levels \cite{Lang17}. In the present work, we do not focus on a particular physiological aspect but rather adopt a global (bio)mechanical point of view. Our main goal is to show that tools derived from mechanics may lead to conserved measurable parameters that the brain can exploit to plan and execute actions instead of relying on an estimate of gravity acquired through a noisy and distributed process. 

Finding such conserved quantities demands to know the underlying dynamics, which is not an easy task in human motion where a same -- even simple -- action requires sometimes very different motor commands. For instance, consider reaching for a cup of coffee on the breakfast table or in an aircraft subject to turbulences, or drinking while seated or while walking. Planning efficient actions is challenging for the brain. It is therefore natural that the central nervous system relies on constants in this jungle of variability. Previous research demonstrated that some classes of actions result from an optimization process in which movements features are taken into account, such as minimizing jerk \cite{viviani95}, metabolic cost \cite{alexander97,berniker} or maintaining a nearly constant mechanical energy in level walking \cite{cavagna}. Here we study rhythmic motion in time-changing gravity as a peculiar case of time-dependent dynamical system with bounded motion. The most powerful tools known so far to study such systems are adiabatic invariants \cite{L&L,henrard,Jose}. They have been applied in a wide range of applications such as plasma physics \cite{notte93} and cosmology \cite{cotsakis98}. In biomechanics, several studies have showed the invariance of the action variable in time-independent conditions \cite{TUR,kugler:1990,turvey:1996,kadar:1993}. In a previous work, we went a step further and proposed to use adiabatic invariants to study the changes in the motion of upper arm rhythmic movements about the elbow at free pace and amplitude in a centrifuge where the perceived gravity's intensity changed stepwise from 1 to 3 $g$ and from 3 back to 1 $g$ \cite{PhysRevE.102.062403}. The direction of $\vec g$ was unchanged. It appeared that the behaviour of the  adiabatic invariant $I = \frac{1}{2\pi}\oint_{\Gamma} P\, dQ$ computed from the latter one-degree-of-freedom motion was compatible with a theoretically predicted linear increase with $g$ \cite{Boulanger:2018tue}. 

An obvious direction to generalize the framework of Ref. \cite{PhysRevE.102.062403} is that of motions involving more than 1 degree of freedom. We first show in Sec. \ref{sec:model} that a linear link between the adiabatic invariant and $g$ is expected for any potential energy, only assuming a separable dynamics. Our model is then applied to analyse the data of Ref. \cite{white08} in Sec. \ref{sec3}. In this last work, participants were asked to continually perform an $\infty-$shaped trajectory during a parabolic flight. The three-dimensional kinematics of the hand has been recorded and adiabatic invariants can be computed from it. Gravity during the parabolas varied between 0 and $1.8$ $g$. We present our results in Sec. \ref{sec:res} and discuss them in Sec. \ref{sec:conclu}.

\section{The model}\label{sec:model}
\subsection{Adiabatic invariants in variable gravity}

We assume that human voluntary rhythmic motion in variable gravity $g(t)$ may be seen as a dynamical system with bounded motion in phase space, described by a separable Hamiltonian $H(I_\alpha, \theta^\alpha,\lambda(t))=\sum^D_{\alpha=1}H_j(I_\alpha, \theta^\alpha,\lambda(t))$. The Hamiltonian depends on action-angle variables $I_\alpha$ and $\theta^\alpha$ respectively and on a time-dependent function $\lambda(t)$, accounting for the modifications induced by $g(t)$. The various ingredients underlying the latter assumption deserve further comments. First, Hamiltonian dynamics being the most powerful formulation of classical Mechanics, it is rather natural to adopt a Hamiltonian approach. 
This being said, not every dynamics is Hamiltonian: A sufficient criterion for a Hamiltonian to exist is that the total time-derivative applied to the Poisson bracket of two functions defined on phase-space obeys the Leibniz rule \cite[Chapter 5]{Jose}. 
The latter criterion cannot a priori be  checked from our experimental setup: We will a posteriori confirm 
 that the observed phase-space trajectories are not incompatible with a Hamiltonian dynamics. Second, the dynamics has a priori no  reason to be separable. The main interest we have in 
 assuming the separability is that it allows a clear separation between vertical and horizontal 
 directions (with respect to $\vec g$), the dynamics in the vertical direction being intuitively the 
 most strongly impacted by the variations of $g\,$. Again, the separability hypothesis will only be checked a posteriori by observing the phase-space trajectories, see next Section.

It can then be shown that \cite[Eqs (50,10)-(50,11)]{L&L}
\begin{subequations}
\begin{eqnarray}
	\dot I_\alpha&=&-\frac{\partial H}{\partial \theta^\alpha}=-\left( \frac{\partial \Lambda}{\partial \theta^\alpha}\right)_{I_\alpha,\lambda}\dot \lambda\ ,  \label{eom1}\\
	\dot \theta^\alpha&=&\frac{\partial H}{\partial I_\alpha}=\omega_\alpha+\left( \frac{\partial \Lambda}{\partial I_\alpha}\right)_{\theta^\alpha,\lambda}\dot \lambda\ ,
\end{eqnarray}
\end{subequations}
where the partial derivatives have to be computed while keeping constant the indexed variables and where $\omega_\alpha$ are the motion's frequencies. The function $\Lambda$ is the action of the system; it is sufficient for our purpose to state that it is a periodic function of the angle variables. Hence, according to \cite{L&L}, $\Lambda=\sum^{+\infty}_{\ell_1=-\infty}\dots\sum^{+\infty}_{\ell_D=-\infty} {\rm e}^{i\vec \ell\cdot \vec \theta}\Lambda_{\vec \ell}\ $ with $\Lambda_{\vec \ell}\in\mathbb{C}$, $\vec \ell=(\ell_1,\dots,\ell_D)\in\mathbb{Z}^D$ and
\begin{equation}
	\frac{\partial \Lambda}{\partial \theta^\alpha}= \sum^{+\infty}_{\ell_1=-\infty}\dots\sum^{+\infty}_{\ell_D=-\infty}i\ell_\alpha  {\rm e}^{i\vec\ell\cdot \vec \theta}\Lambda_{\vec\ell}\ .
\end{equation}

We moreover assume that $\lambda=\lambda_0+\epsilon g(t)$ with $\epsilon g(t)\ll \lambda_0$, \textit{i.e.} that the modifications induced by variable gravity may be computed at first-order in $\epsilon$. Equation (\ref{eom1}) therefore leads to  
\begin{equation}
	\dot I_\alpha=- \epsilon\, \dot g \sum^{+\infty}_{\ell_1=-\infty}\dots\sum^{+\infty}_{\ell_D=-\infty}i\ell_\alpha  {\rm e}^{i\vec\ell\cdot \vec \theta}\Lambda_{\vec\ell}\ ,  
\end{equation}
or 
\begin{equation}\label{eq1}
	\frac{d I_\alpha}{d g}=- \epsilon\, \sum^{+\infty}_{\ell_1=-\infty}\dots\sum^{+\infty}_{\ell_D=-\infty}i\ell_\alpha  {\rm e}^{i\vec\ell\cdot \vec \theta}\Lambda_{\vec\ell}\ .
\end{equation}

Let us define the times $t_n$ such that the values $\theta^\alpha(t_n)$ are all equal (modulo $2\pi$). The existence of $t_n$ is guaranteed for periodic dynamics such as the one that we consider here. Then it can be said that 
\begin{equation}\label{model}
\left. \frac{d I_\alpha}{d g}\right|_{t=t_n}=I_1 \Rightarrow I_\alpha(t_n)=I_{\alpha;0}+I_{\alpha;1}\, g(t_n),
\end{equation}
with $I_{\alpha;0}$ and $I_{\alpha;1}$ two real numbers such that $\vert I_{\alpha;1}/I_{\alpha;0}\vert \ll 1$. Equation (\ref{model}) defines our model: The adiabatic invariant is expected to behave linearly in $g$ when computed at a given position in the consecutive cycles performed. 

\subsection{Definition of $I_\alpha$}
The action variables are defined from positions $(Q^\alpha)$ and momenta $(P_\alpha)$ degrees of freedom as follows:
\begin{equation}\label{Id0}
	I_\alpha = \frac{1}{2\pi}\oint_{\Gamma_\alpha} 
	P_\alpha\, dQ^\alpha\ ,
\end{equation}
where $\Gamma_\alpha$ is the projection of the bounded trajectory in the plane $(Q^\alpha,P_\alpha)$ 
for fixed $\alpha\,$. Note that, with a kinetic energy of the standard form $E_c\sim \sum^D_{\alpha=1}\dot Q^{\alpha\, 2}$, one is led to a form for the adiabatic invariant which is straightforward to compute: 
\begin{equation}\label{Id1}
	I_\alpha(t) \sim \int^{t+T}_{t}\dot Q^{\alpha\, 2}(u)\, du\, ,
\end{equation}
with $T$ the period of the phase-space cycle $\Gamma_\alpha$ starting at $t$. 

This last equation provides a way to compute the adiabatic invariant from experimental data provided $Q^{\alpha}(t)$ is known, which is not so obvious since a mathematical description of voluntary human motion may involve higher derivative dynamics, see \textit{e.g.} \cite{Nelson83,Hogan84,Hagler2015}. Two cases should therefore be considered. First, the motion's dynamics does not involve higher derivative terms. In this case $Q^\alpha$ may directly be identified to, say, one anatomical landmark's trajectory $x^\alpha(t)$, and $P_\alpha\sim \dot x^\alpha$. Second, the motion's dynamics is a higher-derivative one. Then $Q^\alpha$ and $P_\alpha$ can, in principle, be computed from $x^\alpha(t)$ but their definition is more involved. We refer the interested reader to the case of Pais-Uhlenbeck oscillator \cite{Pais:1950za}, that is a higher-derivative generalization of standard harmonic oscillator for which adiabatic invariants can be analytically computed \cite{Boulanger:2018tue}.

\section{The experiment}\label{sec3}
\subsection{Parabolic flights}

\begin{figure}
	\includegraphics[width=\columnwidth]{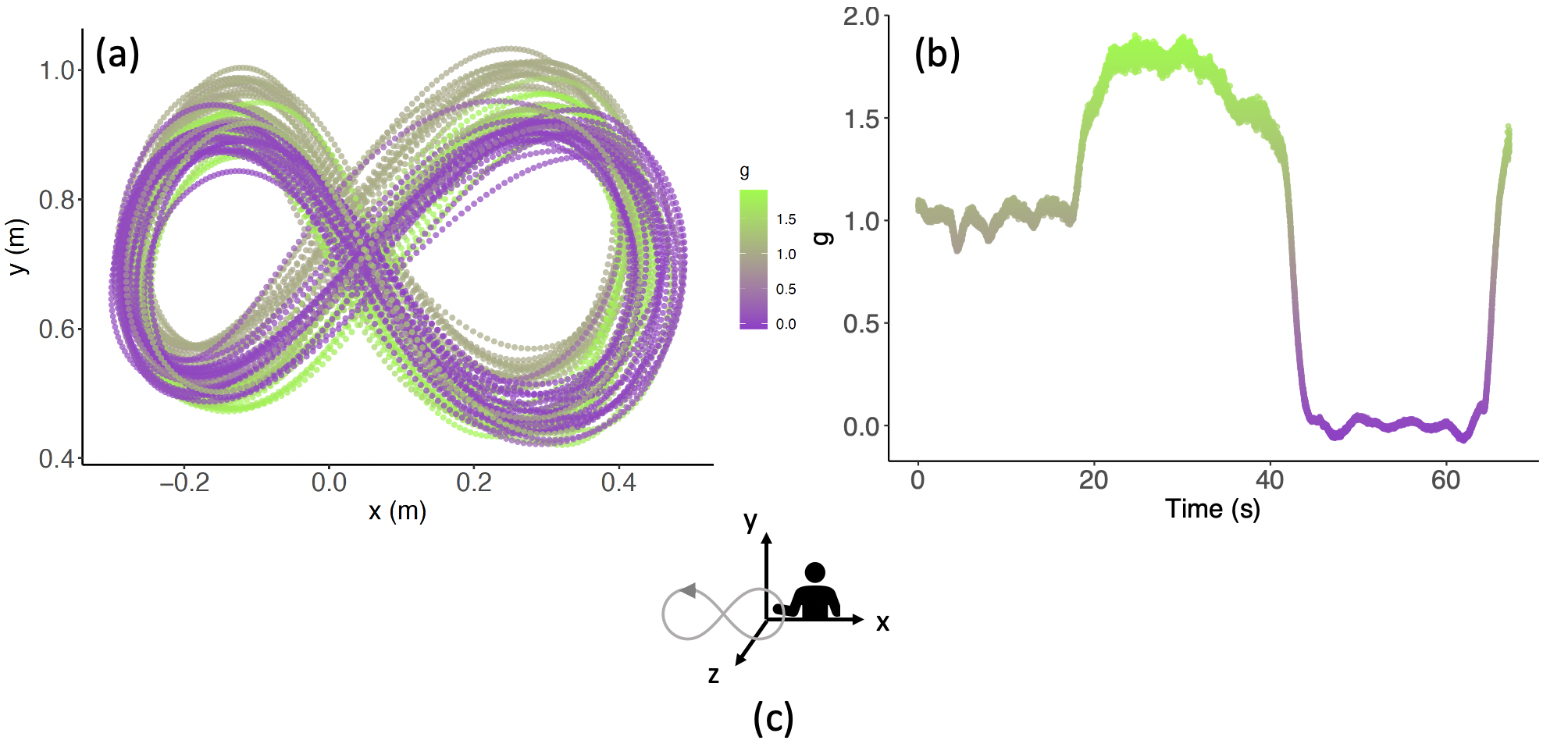} 
	\caption{(a) Typical plot of the $\infty-$shaped motion in frontal plane $(x,y)$ during one parabola. A participant in FREE condition has been chosen. (b) $g(t)$ profile during the same parabola. (c) The Cartesian frame is displayed.  
	}\label{fig:typical}
\end{figure}

During parabolic flights, participants were asked to continually perform an $\infty$-shaped trajectory oriented crosswise to the body around two virtual obstacles situated 3 m in front of them. An optoelectronic device (OptoTrak 3020 system, Northern
Digital, Waterloo, Ontario, Canada) recorded the position of three
infrared LEDs placed on the object with a resolution of 0.1 mm. A three-dimensional
accelerometer fixed on the floor of the aircraft recorded its acceleration. The two synchronized acquisition systems recorded parameters at a sampling rate of 200 Hz. During a parabola, the aircraft performs a series of manoeuvres to allow for changes of effective gravity. This allows to run experiments at 0 (microgravity), 1 and approximately 1.8 $g$ (hypergravity), albeit for a short time. The micro and hyper gravity phases last around 20 s with transition periods shorter than 5 s. Typical plots of the motion performed and of the $g(t)$-profile are shown in Fig. \ref{fig:typical}. The cartesian frame we use is also displayed.

Participants executed $\infty-$shaped movements in two conditions. In the free condition (FREE), the motion was self-paced. Four participants, totalling 24 parabolas, performed the motion in FREE condition. In the metronome condition (METRO), participants had to adopt 1.5-second constant pace prompted by a metronome. Seven participants, totalling 42 parabolas, performed the motion in METRO condition. Before starting the parabolic flights, participant's health was assessed by their individual National Centres for Aerospace Medicine as meeting the requirement "Jar Class II" for parabolic flight. No participant reported sensory or motor deficits and they all had normal or corrected-to-normal vision. All participants gave their informed consent to participate in this study and the procedures were approved by the European Space Agency (ESA) Safety Committee and by the local ethics committee. Their motion was recorded during 6 consecutive parabolas. We refer the reader to Ref. \cite{white08} for a more detailed presentation of the experiment.

\begin{figure}[ht]
	\includegraphics[width=\columnwidth]{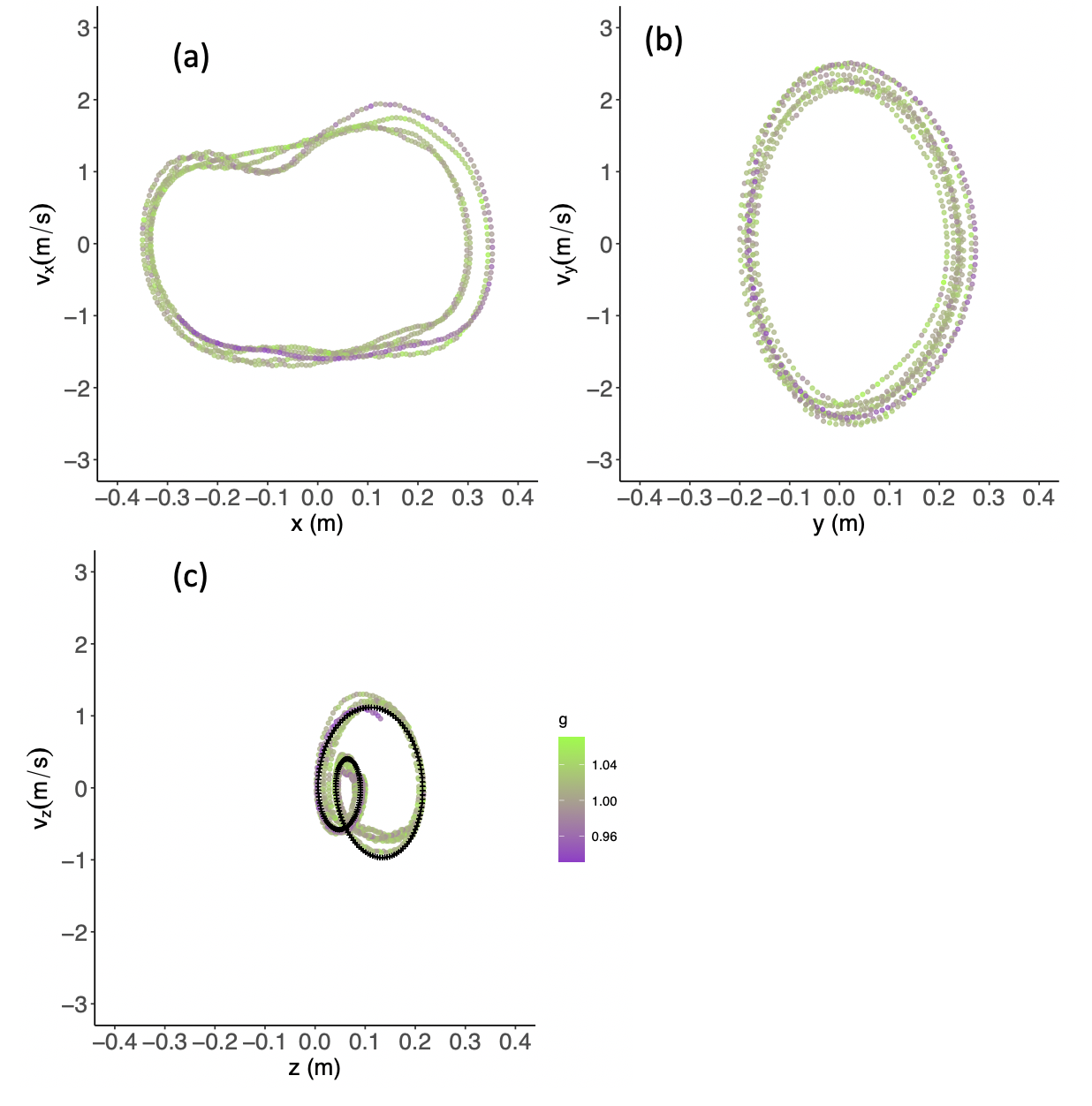} 
	\caption{(a) Typical speed-position plot of the motion in $(x,v_x)$ plane during several cycles, same participant as Fig. \ref{fig:typical}. (b) and (c) Same data in the $(y,v_y)$ and $(z,v_z)$ planes respectively. The crosses show the trajectory obtained from Eq. (\ref{hdt}) with $A_0=0.092$ m, $A_1=0.065$ m, $A_2=0.060$ m, $\omega=2\pi$ rad/s, $\phi_1=0$ rad and $\phi_2=-1$ rad.
	}\label{fig:PS}
\end{figure}

\subsection{Phase-space trajectories and action variables}

The speeds $v_\alpha=\dot x^\alpha$ are first computed from the positions $x^\alpha$ recorded by the optoelectronic device through a finite differentiation. Typical speed-position plots are shown in Fig.~\ref{fig:PS}.
 
The $x$ and $y$ directions show (quasi)-periodic trajectories of elliptic type, compatible with a standard Hamiltonian dynamics. Hence we proceed as follows to compute the action variables. First, we identify $Q^\alpha$ to $x^\alpha$ and $P_\alpha$ to $\dot Q^\alpha$ -- up to an arbitrary mass scale that is set equal to 1 kg. Second, the beginning and end of each cycle $\Gamma_\alpha$ in phase-space plane $(Q^\alpha,P_\alpha)$ are computed. The end of the cycle starting at $t$ is chosen as the time $t^\star$ which is the smallest time after $t$ at which the euclidean distance between $(Q^\alpha(t^\star), P_\alpha(t^\star))$ and $(Q^\alpha(t), P_\alpha(t))$ is minimal. Once $t^\star$ is identified, the adiabatic invariant is computed by quadrature from Eq. (\ref{eq1}): $I_\alpha(t) = \int^{t+t^\star}_{t}\dot Q^{\alpha\, 2}(u)\, du\, $. Then, to apply our model, only adiabatic invariants corresponding to a given value of the angle variable have to be collected. We only consider the instants at which $P_\alpha=0$ and $Q^\alpha$ was maximal since they are easily identified.

As can be seen in Fig. \ref{fig:PS}, the trajectory in $(z,v_z)$ plane intersects itself during one cycle. The underlying dynamic is therefore called \emph{non-autonomous}, in the theory of dynamical systems.Participant's motion in the $z-$direction actually contains two distinct frequencies. One is the whole $\infty$-shaped movement's pulsation, say $\omega$, and the other is a forward-backward oscillation at $2\omega$. A trajectory of the form 
	\begin{equation}\label{hdt}
		z(t)=A_0+\sum_{j=1,2}A_j\sin(j \omega t+\phi_j)
	\end{equation} 
has the qualitative features of what is observed in the $(z,v_z)$ for appropriate values of the real constants $\omega$, $A_i$ and $\phi_i$.
	
Several effective models may produce trajectories such as (\ref{hdt}). (1) An oscillator with pulsation $\omega$ plus an external, time-dependent, periodic force with pulsation $2\omega$. A textbook example is the Duffing oscillator. The complete set of solutions of a forced, non-harmonic oscillator is unknown a priori but some special solutions are known that perfectly match the observed motion. (2) A system of coupled harmonic oscillators oscillating near its equilibrium position provided that the frequencies of some normal modes are equal to $\omega$ and $2\omega$. The dynamics of the various joints of the arm may be approximated by such a system. (3) A second-order Pais-Uhlenbeck harmonic oscillator. As shown in Appendix \ref{app}, an autonomous higher-derivative oscillator may actually mimic the dynamics of one peculiar degree of freedom in a system of coupled oscillators. An obvious advantage in resorting to Pais-Uhlenbeck harmonic oscillators is that the effective dynamics in the $z-$direction would be autonomous and that adiabatic invariants are well-defined once phase-space is properly built from the position degree of freedom and its time derivatives \cite{Boulanger:2018tue}. Hence, our model can in principle be adapted to the $z-$direction. However, such higher-derivative adiabatic invariants involve not only $\dot z$ but at least $\ddot z$ and $\dddot{z}$ \cite{Boulanger:2018tue}. The experimental precision reached in the measurement of $z$ does not allow for a reliable computation of those higher derivatives from our experimental data and we chose not to make further computations as far as the $(z,v_z)$ plane is concerned.

\section{Action variables in terms of gravity: The results}\label{sec:res}

We have linearly fitted $I_\alpha$  versus $g$ for each available parabola in order to check whether Model (\ref{model}) is observed at an individual level or not. The variable $g$ refers to the average value of $g(t)$ within the considered cycle $\Gamma_\alpha$. The parameters of the fit are $I_{1;\alpha}$ (slope),  $r_\alpha$ (Pearson's correlation coefficient) and $I_{0;\alpha}$ (intercept). A two-way ANOVA may be performed on the parameters of the fit with factors condition (FREE or METRO) and parabola number (1 to 6). The latter factor is introduced to check whether a learning effect is present or not during the consecutive parabolas a given participant has experienced.  The ANOVA was performed using SigmaPlot software (v.11.0, Systat Software, San Jose, CA, United States of America), with significance level 0.05. It appears that no significant effect of the parabola number can be found in the fit parameters which means that the model parameters are stable over time, with values set from the outset. Interactions between condition and parabola number are not significant either. However, the condition has a significant impact on $I_{0;\alpha}$, $I_{1;y}$  and $r_y$, as shown in Table \ref{table:fit1}. 

\begin{figure}
	\includegraphics[width=0.6\columnwidth]{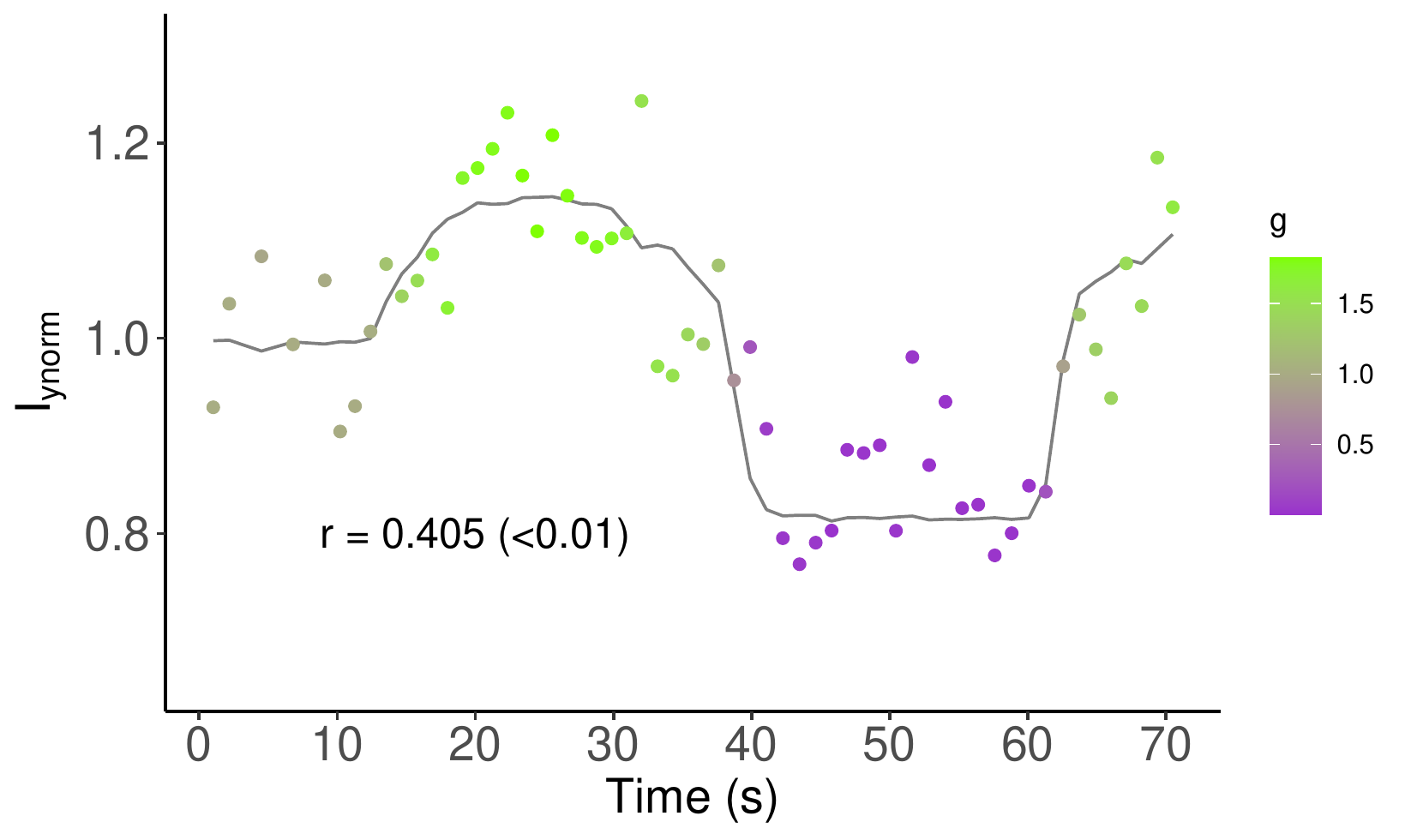} 
	\caption{Adiabatic invariant $I_y$ versus $g$ computed from experimental data in the FREE condition for the same participant as in Fig. \ref{fig:typical} (points), compared to the best linear fit of the form (\ref{model}). Pearson's correlation coefficients is also indicated. $I_y$ has been normalized so that its average value is $1$ at $1g$. 
	}\label{fig:Ibest}
\end{figure}

From Table \ref{table:fit1}, the following global features of participant's motion can be deduced. (a) The action variable $I_x$ does not show a well-defined linear behaviour versus $g$: both the slopes and the Pearson's correlation coefficients are comparable with 0. The action variable $I_x$ can be considered as constant with $g$, although it is significantly lower in METRO condition than in the FREE condition. Let us note that the $x-$direction is orthogonal to gravity while the $y-$direction is aligned with gravity (see frame in Fig. 1). This observation may intuitively explain why there is no trend versus $g$ for the $x-$dynamics. (b) However, the trend of $I_y$ vs $g$ is compatible with model (\ref{model}) for positive $I_{1;y}$ well smaller than the intercept $I_{0;y}$. It is coherent with our initial assumption to work at first-order in $g$. Furthermore, the slope and the intercept are significantly lower in METRO condition than in the FREE one. (c) The clearest linear trend is observed for the action variable $I_y$ in the FREE condition. An example of linear fit is displayed in Fig. \ref{fig:Ibest}.

\begin{table}
	\begin{tabular}{ll|rrr}
		Direction& Condition & $I_{1;\alpha}$ (kg.m) & $r_\alpha$ & $I_{0;\alpha}$ (J.s)\\
		\hline
		$x$ & FREE & $[-0.014,0.005]$ & $[-0.150,0.082]$ & $[0.327,0.371]$ \\
		    & METRO & $[-0.010,0.004]$ & $[-0.182,0.109]$ & $[0.177,0.205]$\\
		    & $p$ & 0.884 & 0.969 & $ <0.001$ \\
		\hline
		$y$ & FREE & $[0.017,0.033]$ & $[0.291,0.518]$ & $[0.187,0.296]$ \\
			& METRO & $[0.005,0.14]$ & $[0.117,0.332]$ & $[0.106,0.120]$\\
 			& $p$ & $ <0.001$  & $ 0.014$  & $ <0.001$  \\
	\end{tabular}
	\caption{95 \% confidence intervals for the slopes $I_{1;\alpha}$, Person's correlation coefficients $r_\alpha$ and intercepts $I_{0;\alpha}$ obtained through the fit (\ref{model}) of the computed $I_\alpha$ vs $g$ in each parabola for all conditions. The $p-$ values of the ANOVA for the effect of condition are also given. }
	\label{table:fit1}
\end{table}

\begin{figure}
	\includegraphics[width=0.6\columnwidth]{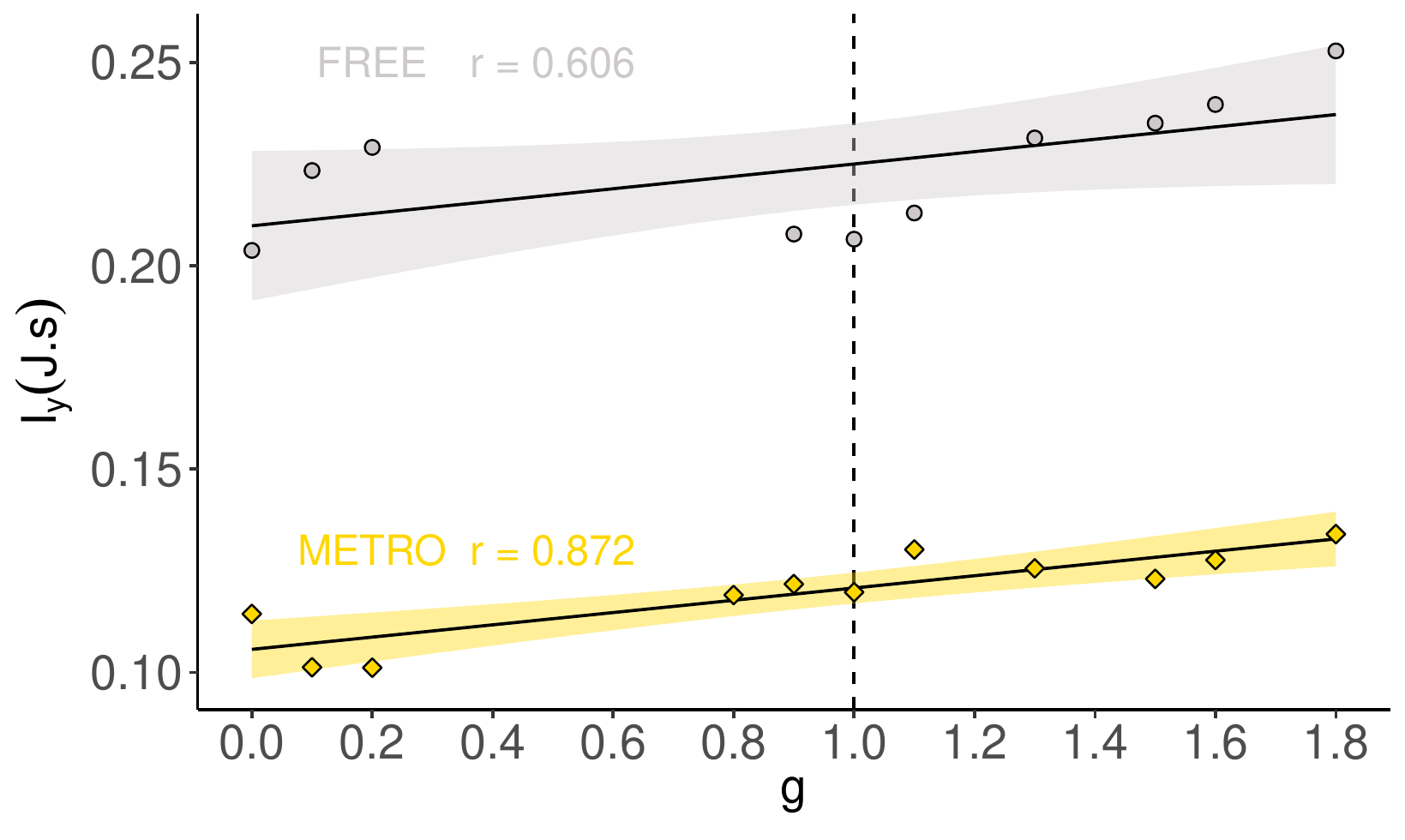} 
	\caption{Adiabatic invariant $I_y$ versus $g$ computed from experimental data in the FREE (grey points) and METRO (yellow points) conditions. A linear fit is given (solid lines) with its 95\% confidence interval (colored bands) in each condition. Pearson's correlation coefficients are also indicated. The 1-$g$ bin is marked with a vertical dashed line. 
	}\label{fig:Iy}
\end{figure}

The global trend of $I_y$ vs $g$ can be observed by averaging $I_y$ over participants by condition (FREE and METRO) and by gravity condition, \textit{i.e.} by gathering computed adiabatic invariants into bins of 0.1 $g$, ranging from 0 to 1.8 $g$. Only the bins containing more than 10 points were finally kept. This threshold is arbitrary but avoids almost empty bins in the fast transition regions between 0 and 1 $g$ and between 1 and 1.6 $g$. The results of this analysis are displayed in Fig. \ref{fig:Iy}. The observed trends are compatible with $I_y^{FREE}=0.210+0.015\, g$ and  $I_y^{METRO}=0.106+0.015\, g$; an ANCOVA further shows that the slopes are not significantly different with condition ($p=0.994$), while the intercepts significantly depend on condition ($p<0.001$). Finally, it is also worth highlighting the fact that some bins (e.g. $[1.3;1.4[$) capture values of gravity in the ascending but also descending parts of the parabolic profile. 

\section{Discussion of the results}\label{sec:conclu}

The trajectory in the $z-$direction contains two distinct frequencies leading to intersecting trajectories in the $(z,v_z)$ plane, see Fig. \ref{fig:PS}. Effective dynamics in this plane therefore cannot be described by a time-independent standard dynamics: Either higher-derivatives or time-dependent forces have to be included. We believe that the appearance of such features could be explained by shoulder biomechanical constraints. The main shoulder movements required to execute the $\infty-$shaped movement are abduction and adduction. During shoulder abduction-adduction movements, rotations are usually observed \cite{assi}. It is well known that external rotation of the shoulder is adopted during abduction to clear the major tubercule of humerus from beneath acromion for preventing impingement \cite{peat,hurov}. The movement strategy spontaneously chosen by participants is therefore not  located on a single plane with constant $z$, leading to the observed nontrivial pattern.As previously said, the current experimental accuracy along the $z-$axis does not allow for a more detailed study of a potential higher-derivative effective dynamics. Note that it has already been successfully conjectured that a higher-derivative action principle such as $S=\int \dddot x^2\, dt$ -- a jerk-based cost function -- may constrain non-rhythmic voluntary human motion \cite{Hogan84}. However, such an action principle does not lead to periodic solutions, that is why a Pais-Uhlenbeck oscillator seems more relevant to us. 
In the motion we observe, the two frequencies have an integer ratio, therefore stability of the motion is not guaranteed in such a resonant case \cite{Boulanger:2018tue}. It has recently been understood that extra interaction terms may stabilize periodic solutions of resonant higher-derivative oscillators \cite{Kaparulin:2020rqz}: We hope to investigate the applicability of such models to human rhythmic motion in future works.

By definition, the adiabatic invariant in $x$ and $y-$directions is proportional to 
\begin{equation}\label{I_def1}
	I_\alpha\sim T \left\langle E_{c,\alpha} \right\rangle ,
\end{equation}
where $T$ is a given cycle duration in the vertical direction -- the duration of the cycle starting at the same time is twice that value in the $x-$direction --, and where $\left\langle E_{c,\alpha} \right\rangle$ is the averaged kinetic energy on the considered cycle. The protocol of \cite{white08} is such that $T^{FREE}<T^{METRO}$: The pace imposed by the metronome was chosen to be slower than participants' spontaneously chosen paces. Since, at given $g$, the adiabatic invariant in FREE condition is always larger than in METRO condition, it can be concluded that $\left\langle E_{c,\alpha}^{FREE} \right\rangle > \left\langle E_{c,\alpha}^{METRO} \right\rangle$. The smaller kinetic energy in METRO condition thus follows from the fact that participants have to move slower than in FREE condition in order to follow metronome's pace. Note that $I_{1;y}^{METRO}<I_{1;y}^{FREE}$: The extra constraint imposed by the metronome actually prevents the participant from optimally adapting his/her motion when $g$ is changing, assuming that the optimal motor strategy is reached in FREE condition. It is also known that $T$ is a decreasing function of $g$ in either METRO or FREE conditions \cite{white08}. Since $I_{1;y}>0$, $\left\langle E_{c,y} \right\rangle $ has to be an increasing function of $g$: The participant's arm moves with higher typical vertical speed at higher values of $g$. 

Definition (\ref{I_def1}) explicitly makes appear the links between adiabatic invariant and kinetic energy. Another interpretation of the adiabatic invariant, focusing on external forces, is relevant to clarify the influence of gravity on it. To this aim, the virial theorem may be used to state that 
\begin{equation}\label{I_av}
	I_\alpha\sim T \left\langle F_{\alpha}x^\alpha  \right\rangle,
\end{equation}
where $F_\alpha$ is an external force acting on the point-like object whose trajectory is $x^\alpha$. The latter force should involve muscular forces as well as gravity. One can reasonably assume that $F_\alpha=F_{0\alpha}+F_{1\alpha} g$, in coherence with the previously found linear trend of $I_\alpha$ vs $g$. A priori, $F_{1y}\gg F_{1x}$ since gravity's influence should mostly concern the vertical direction. Changes induced by $F_{1x}$ are probably unnoticeable up to our current experimental precision. 

Around $1g$, $I_y$ is lower than expected from the 95\% confidence interval of the linear fit. In that familiar environment, participants ``know" the most economic strategy when they are allowed to move freely in Earth's gravity. In METRO condition, that drop in $I_y$ is not observable: The extra constraint imposed by the metronome does not allow participants to follow that optimal strategy. Furthermore, our results also reveal that microgravity is a special case. While the linear fit holds true for the whole explored gravitational values ($[0 g ; 1.8 g]$), there is a significant gap between 0.3 $g$ and 0.7 $g$ in our data. Hypogravity values are not explored. 
Our study again reveals that 0 $g$ acts as a singular value for the brain \cite{white08}. Finally, adiabatic invariants do not behave like parameters measured in most motor control investigations. Indeed, while errors in reaching movements perturbed by force fields require tens of trials to vanish \cite{shad94}, safety margins in object manipulation in altered gravity need an exposure to 6 parabolas to decrease to normal values \cite{augu03} and some behaviours in conflicting force-fields or visuomotor rotations do even not adapt at all \cite{cothros}, adiabatic invariants seem to be set to their nominal values from the outset.

\section{Concluding comments}

To conclude this work, it is worth linking them to well-known frameworks in motor control.  

Participants have many more kinematic degrees of freedom than necessary to fulfill the demanded task, \textit{i.e.} the $\infty-$shaped movement. The coordination of kinematically redundant systems was formulated by Bernstein as the degrees of freedom problem \cite{Bernstein}. The main difficulty of Bernstein's problem is that the nervous system must conciliate two apparently conflicting abilities: (1) the realization of a movement from the choice of one among an infinite number of motor patterns; (2) the absence of univocal relationship between the movement realized and motor patterns used, known as motor equivalence. Although it remains unclear as to whether and how the brain can estimate adiabatic invariants, such quantities puts constraints on the allowed strategies, \textit{i.e.} strategies keeping $I_\alpha$ invariant at constant $g$. In this picture, an increase (decrease) in $I_\alpha(g)$ may be related to an increase (decrease) of the allowed motor patterns. 

When subjects are free to point to a target, they automatically scale movement duration with movement amplitude and choose a trade off between movement speed and accuracy to touch the target. It is known as Fitts’s law \cite{Fitts}. The adiabatic invariant is the area of a closed trajectory in phase space: $I_\alpha\sim A_\alpha\, v^{{\rm max}}_\alpha$, with $A_\alpha$ and $v^{{\rm max}}_\alpha$ the amplitude and maximal speed of the movement in the direction $\alpha$ respectively. Its invariance at given $g$ implies that, if maximal speed increases (decreases), amplitude decreases (increases). In our experiment, the maximal speed is an obvious measure of movement's speed, and the amplitude can be seen as an index of precision. Indeed,the instruction given to the participant is to avoid 2 targets by turning around. Thus, if amplitude decreases (increases), the participant increases (decreases) the chances of hitting the target, and he/she is less (more) precise. The adiabatic invariant can then be seen as an explicit realization of the speed-accuracy-trade off scenario. The modification of its value with $g$ actually changes the acceptable values of maximal speed and amplitude involved in this trade off.

In summary, our results indicate that adiabatic invariants deserve a particular attention in biomechanical approaches of human motion. They are indeed able to capture one individual's reaction to time-dependent external conditions, even in extreme cases such as variable gravity. Adiabatic invariants seem very robust in this context. Further studies are now needed to clarify the links between adiabatic invariant theory and celebrated motor control paradigms such as speed-accuracy-trade off.

\medskip

\textit{Acknowledgements}
This research was supported by grants from Prodex and IAP; Belgian
Federal Office for Scientific, Technical, and Cultural Affairs; Fonds Sp\'ecial de
Recherche; 
and Canadian Space
Agency Contract 9F007-033026. 

\appendix*

\section{Higher derivative dynamics and rhythmic motion}\label{app}

Let us consider a system with $N$ degrees of freedom $x^\alpha$ described by the Lagrangian 
\begin{align}
	L=\frac{1}{2}g_{\alpha\beta}(x)\dot x^\alpha\dot x^\beta-U(x^\gamma)
	\label{lagrangian}
\end{align}
with $g_{\alpha\beta}$ the components of a real, symmetric and positive-definite matrix 
$G$ that we call the kinetic matrix. Note that it is not necessarily constant 
and may depend on the dynamical variables.
If necessary after a translation of the origin of the coordinates, 
we may assume that $x^\alpha=0$ ($\forall ~\alpha$) is an equilibrium 
position: $\left. \frac{\partial U}{\partial x^\alpha}\right|_{x^\gamma=0}=0\,$. 
Such a Lagrangian may model the motion of several joints, the potential energy $U$ being an a priori complicated function of the degrees of freedom. 
If only small oscillations around equilibrium position are considered, the equations of motion read 
$\ddot x^\alpha+\gamma^{\alpha\delta}U_{\delta\beta} x^\beta=0\,$, 
with $\gamma^{\alpha\beta}$ the components of the inverse of the matrix 
$G_{0}$ of components $\gamma_{\alpha\beta}:=g_{\alpha\beta}(0)\,$.
In other words, one has $\gamma^{\alpha\delta}\gamma_{\delta\beta}=\delta^\alpha_\beta\,$. 
One defines the potential matrix $U$ with components
\begin{equation}
	U_{\alpha\beta}=\left. \frac{\partial^2U}{\partial x^\alpha \partial x^\beta}\right|_{x^\delta=0}.
\end{equation}

Solving the eigenequation $V^\alpha{}_\beta\,\xi^\beta_a=\lambda_a\,\xi^\alpha_a$ 
for the matrix $V=G_{0}^{-1}U$ with components 
$V^\alpha{}_\beta=\gamma^{\alpha\delta}\,U_{\delta\beta}\,$, with $a=1,\dots,N\,$, 
allows to solve the equations of motion in terms of the normal coordinates $Q^a(t)$: 
\begin{equation}
x^\alpha=\xi^\alpha_a\,Q^a(t)\ \qquad \mbox{with} \qquad 
	\ddot Q^a=-\lambda_a \,Q^a
\end{equation}
and without any summation over the index $a\,$.
Therefore, for the given dynamical system described by the variables $x^{\alpha}$,  
any small oscillatory motion about the minimum of the potential in configuration space 
can therefore be decomposed as a linear combination of elementary oscillations along 
the normal modes, each one at the frequency 
$\nu_{a}=\omega_{a}/2\pi\,$, where $\lambda_{a}=\omega_{a}^{2}\,$. In particular, 
for appropriate initial conditions it is possible to only excite the normal mode $Q^{a}(t)$ 
for a given  value of the index $a\,$. The dynamical system as a whole will then oscillate 
at the single frequency $\nu_{a}\,$, without exciting the modes $Q^{b}(t)$ with $b\neq a\,$.
The interested reader 
may find a detailed discussion about small oscillations around an equilibrium position in 
\cite[Chapter 5]{L&L}, or in \cite[Part 2, Chapter 5]{Arnold} for a more precise mathematical 
formulation. 

Note that, from the datum of the normal modes $Q$'s with their frequencies $\nu$'s, 
one can go back and access the information contained in the kinetic and potential
matrices $G\,$ and $U\,$. This is because the normal modes are orthogonal with 
respect to the metric $G\,$, and using the latter metric together with the eigenvalues
of $V:=G^{-1}U\,$ gives $U\,$ up to a reordering of the dynamical variables $x^{\alpha}\,$.

In a human rhythmic motion, if the participant is asked to perform a periodic motion, 
say with the forearm, 
one observes that the projection of the motion of the hand along the three spatial 
directions gives rise to a very small set of frequencies that are all integer multiples of  
a fundamental one. In this description, we neglect the quasi-periodic motion of the 
forearm due to physiological noise.
Of course, the forearm is a very complicated system with dozens of components 
linked in a complicated fashion, giving rise to a configuration space $\mathbb{Q}$
of very large dimension $N\,$. In principle (but not in practice), it is possible to describe 
it by a Lagrangian of the form \eqref{lagrangian} and there will $N$ normal 
modes $Q^{a}$'s with possible degeneracies in the frequencies. 
The observed motion of the forearm of the participants is, instead, 
very simple and degenerate. 

Instead of trying to find the realistic Lagrangian description \eqref{lagrangian} 
of the forearm, from the sole observation of a very 
limited set of forced periodic motions with distinct frequencies $\omega_{a}$'s,  
motions that we view as analogous to the distinct normal modes of a dynamical system, 
we propose an 
effective model whose purpose it to reproduce those ``normal modes'' without any 
diagonalisation of any potential matrix $V\,$. 
The operator
\begin{equation}
	F=\Pi^n_{a=1}\left(1+\frac{1}{\omega^2_a}\frac{d^2}{dt^2}\right)
\end{equation} 
is such that $F x^\alpha=0\,$ \emph{for all} $\,\alpha$ because $FQ^a=0$ for all $a\,$, by construction.
Here, by an abuse of notation we have denoted by $Q^{a}(t)$ the $n\ll N$ simple 
and pure harmonic modes observed in the participant's motion. 
The latter describes the motion in a configuration space of very large dimension $N\,$ whereas we
effectively reduce the dynamics to a configuration space of dimension $n$ way smaller than $N\,$.    
Therefore, in our effective description of the motion based on a specific set of harmonic oscillations 
observed in the forearm's motion, every single dynamical variable $x^\alpha$ for $\alpha$ fixed 
obeys the equation of motion 
of a Pais-Uhlenbeck oscillator whose Lagrangian reads 
$L_{P-U}=-\frac{1}{2} \,x^\alpha\, F\, x^\alpha$ \cite{Pais:1950za}. 
If at least two frequencies $\omega_a$ are different, the effective dynamics of a given degree of freedom 
can be mimicked by a particular solution of the equations of motion 
of a higher-derivative harmonic oscillator. 

\bibliographystyle{apsrev}
\bibliography{biblio_parabola}

\begin{thebibliography}{36}
\expandafter\ifx\csname natexlab\endcsname\relax\def\natexlab#1{#1}\fi
\expandafter\ifx\csname bibnamefont\endcsname\relax
  \def\bibnamefont#1{#1}\fi
\expandafter\ifx\csname bibfnamefont\endcsname\relax
  \def\bibfnamefont#1{#1}\fi
\expandafter\ifx\csname citenamefont\endcsname\relax
  \def\citenamefont#1{#1}\fi
\expandafter\ifx\csname url\endcsname\relax
  \def\url#1{\texttt{#1}}\fi
\expandafter\ifx\csname urlprefix\endcsname\relax\def\urlprefix{URL }\fi
\providecommand{\bibinfo}[2]{#2}
\providecommand{\eprint}[2][]{\url{#2}}

\bibitem[{\citenamefont{White et~al.}(2020)\citenamefont{White, Gaveau,
  Bringoux, and Crevecoeur}}]{white20}
\bibinfo{author}{\bibfnamefont{O.}~\bibnamefont{White}},
  \bibinfo{author}{\bibfnamefont{.~J.} \bibnamefont{Gaveau}},
  \bibinfo{author}{\bibfnamefont{L.}~\bibnamefont{Bringoux}}, \bibnamefont{and}
  \bibinfo{author}{\bibfnamefont{F.}~\bibnamefont{Crevecoeur}},
  \bibinfo{journal}{J. Neurophysiol.} \textbf{\bibinfo{volume}{124}},
  \bibinfo{pages}{4} (\bibinfo{year}{2020}).

\bibitem[{\citenamefont{Einstein}(1916)}]{Einstein:1916vd}
\bibinfo{author}{\bibfnamefont{A.}~\bibnamefont{Einstein}},
  \bibinfo{journal}{Annalen Phys.} \textbf{\bibinfo{volume}{49}},
  \bibinfo{pages}{769} (\bibinfo{year}{1916}).

\bibitem[{\citenamefont{Wald}(1984)}]{Wald:106274}
\bibinfo{author}{\bibfnamefont{R.~M.} \bibnamefont{Wald}},
  \emph{\bibinfo{title}{{General relativity}}} (\bibinfo{publisher}{Chicago
  Univ. Press}, \bibinfo{address}{Chicago, IL}, \bibinfo{year}{1984}).

\bibitem[{\citenamefont{Aubert et~al.}(2016)}]{Aubert16}
\bibinfo{author}{\bibfnamefont{A.}~\bibnamefont{Aubert}} \bibnamefont{et~al.},
  \bibinfo{journal}{npj Microgravity} \textbf{\bibinfo{volume}{2}},
  \bibinfo{pages}{16031} (\bibinfo{year}{2016}).

\bibitem[{\citenamefont{White et~al.}(2016)}]{White16}
\bibinfo{author}{\bibfnamefont{O.}~\bibnamefont{White}} \bibnamefont{et~al.},
  \bibinfo{journal}{npj Microgravity} \textbf{\bibinfo{volume}{2}},
  \bibinfo{pages}{16023} (\bibinfo{year}{2016}).

\bibitem[{\citenamefont{Lang et~al.}(2017)\citenamefont{Lang, Van~Loon,
  Bloomfield, Vico, Chopard, Rittweger, Kyparos, Blottner, Vuori, Gerzer
  et~al.}}]{Lang17}
\bibinfo{author}{\bibfnamefont{T.}~\bibnamefont{Lang}},
  \bibinfo{author}{\bibfnamefont{J.}~\bibnamefont{Van~Loon}},
  \bibinfo{author}{\bibfnamefont{S.}~\bibnamefont{Bloomfield}},
  \bibinfo{author}{\bibfnamefont{L.}~\bibnamefont{Vico}},
  \bibinfo{author}{\bibfnamefont{A.}~\bibnamefont{Chopard}},
  \bibinfo{author}{\bibfnamefont{J.}~\bibnamefont{Rittweger}},
  \bibinfo{author}{\bibfnamefont{A.}~\bibnamefont{Kyparos}},
  \bibinfo{author}{\bibfnamefont{D.}~\bibnamefont{Blottner}},
  \bibinfo{author}{\bibfnamefont{I.}~\bibnamefont{Vuori}},
  \bibinfo{author}{\bibfnamefont{R.}~\bibnamefont{Gerzer}},
  \bibnamefont{et~al.}, \bibinfo{journal}{npj Microgravity}
  \textbf{\bibinfo{volume}{3}}, \bibinfo{pages}{8} (\bibinfo{year}{2017}).

\bibitem[{\citenamefont{Viviani and Flash}(1995)}]{viviani95}
\bibinfo{author}{\bibfnamefont{P.}~\bibnamefont{Viviani}} \bibnamefont{and}
  \bibinfo{author}{\bibfnamefont{T.}~\bibnamefont{Flash}},
  \bibinfo{journal}{Journal of Experimental Psychology: Human Perception and
  Performance} \textbf{\bibinfo{volume}{21}}, \bibinfo{pages}{32}
  (\bibinfo{year}{1995}).

\bibitem[{\citenamefont{Alexander}(1997)}]{alexander97}
\bibinfo{author}{\bibfnamefont{R.}~\bibnamefont{Alexander}},
  \bibinfo{journal}{Biol Cybern.} \textbf{\bibinfo{volume}{76}},
  \bibinfo{pages}{97} (\bibinfo{year}{1997}).

\bibitem[{\citenamefont{Berniker et~al.}(2013)\citenamefont{Berniker, O'Brien,
  Kording, and Ahmed}}]{berniker}
\bibinfo{author}{\bibfnamefont{M.}~\bibnamefont{Berniker}},
  \bibinfo{author}{\bibfnamefont{M.~K.} \bibnamefont{O'Brien}},
  \bibinfo{author}{\bibfnamefont{K.~P.} \bibnamefont{Kording}},
  \bibnamefont{and} \bibinfo{author}{\bibfnamefont{A.~A.} \bibnamefont{Ahmed}},
  \bibinfo{journal}{PLOS ONE} \textbf{\bibinfo{volume}{8}}, \bibinfo{pages}{1}
  (\bibinfo{year}{2013}).

\bibitem[{\citenamefont{Cavagna et~al.}(1976)\citenamefont{Cavagna, Thys, and
  Zamboni}}]{cavagna}
\bibinfo{author}{\bibfnamefont{G.}~\bibnamefont{Cavagna}},
  \bibinfo{author}{\bibfnamefont{H.}~\bibnamefont{Thys}}, \bibnamefont{and}
  \bibinfo{author}{\bibfnamefont{A.}~\bibnamefont{Zamboni}},
  \bibinfo{journal}{J Physiol.} \textbf{\bibinfo{volume}{262}},
  \bibinfo{pages}{639} (\bibinfo{year}{1976}).

\bibitem[{\citenamefont{Landau and Lifchitz}(1988)}]{L&L}
\bibinfo{author}{\bibfnamefont{L.}~\bibnamefont{Landau}} \bibnamefont{and}
  \bibinfo{author}{\bibfnamefont{E.}~\bibnamefont{Lifchitz}},
  \emph{\bibinfo{title}{{Physique th\'{e}orique Tome 1 : M\'{e}canique}}}
  (\bibinfo{publisher}{E. MIR}, \bibinfo{address}{Moscow},
  \bibinfo{year}{1988}).

\bibitem[{\citenamefont{Henrard}(1998)}]{henrard}
\bibinfo{author}{\bibfnamefont{J.}~\bibnamefont{Henrard}},
  \emph{\bibinfo{title}{The Adiabatic Invariant in Classical Mechanics}}
  (\bibinfo{publisher}{Dessy}, \bibinfo{year}{1998}), pp.
  \bibinfo{pages}{60--73}.

\bibitem[{\citenamefont{Jose and Saletan}(1998)}]{Jose}
\bibinfo{author}{\bibfnamefont{J.}~\bibnamefont{Jose}} \bibnamefont{and}
  \bibinfo{author}{\bibfnamefont{E.}~\bibnamefont{Saletan}},
  \emph{\bibinfo{title}{Classical dynamics: a contemporary approach}}
  (\bibinfo{publisher}{Cambridge Univ. Press}, \bibinfo{address}{Cambridge},
  \bibinfo{year}{1998}).

\bibitem[{\citenamefont{Notte et~al.}(1993)\citenamefont{Notte, Fajans, Chu,
  and Wurtele}}]{notte93}
\bibinfo{author}{\bibfnamefont{J.}~\bibnamefont{Notte}},
  \bibinfo{author}{\bibfnamefont{J.}~\bibnamefont{Fajans}},
  \bibinfo{author}{\bibfnamefont{R.}~\bibnamefont{Chu}}, \bibnamefont{and}
  \bibinfo{author}{\bibfnamefont{J.~S.} \bibnamefont{Wurtele}},
  \bibinfo{journal}{Phys. Rev. Lett.} \textbf{\bibinfo{volume}{70}},
  \bibinfo{pages}{3900} (\bibinfo{year}{1993}).

\bibitem[{\citenamefont{Cotsakis et~al.}(1998)\citenamefont{Cotsakis, Lemmer,
  and Leach}}]{cotsakis98}
\bibinfo{author}{\bibfnamefont{S.}~\bibnamefont{Cotsakis}},
  \bibinfo{author}{\bibfnamefont{R.~L.} \bibnamefont{Lemmer}},
  \bibnamefont{and} \bibinfo{author}{\bibfnamefont{P.~G.~L.}
  \bibnamefont{Leach}}, \bibinfo{journal}{Phys. Rev. D}
  \textbf{\bibinfo{volume}{57}}, \bibinfo{pages}{4691} (\bibinfo{year}{1998}).

\bibitem[{\citenamefont{Kugler and Turvey}(1987)}]{TUR}
\bibinfo{author}{\bibfnamefont{P.}~\bibnamefont{Kugler}} \bibnamefont{and}
  \bibinfo{author}{\bibfnamefont{M.}~\bibnamefont{Turvey}},
  \emph{\bibinfo{title}{{Information, Natural Law, and the Self-Assembly of
  Rhythmic Movement}}} (\bibinfo{publisher}{London: Routledge},
  \bibinfo{address}{London}, \bibinfo{year}{1987}).

\bibitem[{\citenamefont{Kugler et~al.}(1990)\citenamefont{Kugler, Turvey,
  Schmidt, and Rosenblum}}]{kugler:1990}
\bibinfo{author}{\bibfnamefont{P.}~\bibnamefont{Kugler}},
  \bibinfo{author}{\bibfnamefont{M.}~\bibnamefont{Turvey}},
  \bibinfo{author}{\bibfnamefont{R.}~\bibnamefont{Schmidt}}, \bibnamefont{and}
  \bibinfo{author}{\bibfnamefont{L.}~\bibnamefont{Rosenblum}},
  \bibinfo{journal}{Ecological Psychology} \textbf{\bibinfo{volume}{2}},
  \bibinfo{pages}{151} (\bibinfo{year}{1990}).

\bibitem[{\citenamefont{Turvey et~al.}(1996)\citenamefont{Turvey, Holt, Obusek
  et~al.}}]{turvey:1996}
\bibinfo{author}{\bibfnamefont{M.}~\bibnamefont{Turvey}},
  \bibinfo{author}{\bibfnamefont{K.}~\bibnamefont{Holt}},
  \bibinfo{author}{\bibfnamefont{J.}~\bibnamefont{Obusek}},
  \bibnamefont{et~al.}, \bibinfo{journal}{Biol. Cybern.}
  \textbf{\bibinfo{volume}{74}}, \bibinfo{pages}{107} (\bibinfo{year}{1996}).

\bibitem[{\citenamefont{Kadar et~al.}(1993)\citenamefont{Kadar, Schmidt, and
  Turvey}}]{kadar:1993}
\bibinfo{author}{\bibfnamefont{E.}~\bibnamefont{Kadar}},
  \bibinfo{author}{\bibfnamefont{R.}~\bibnamefont{Schmidt}}, \bibnamefont{and}
  \bibinfo{author}{\bibfnamefont{M.}~\bibnamefont{Turvey}},
  \bibinfo{journal}{Biol. Cybern.} \textbf{\bibinfo{volume}{68}},
  \bibinfo{pages}{421} (\bibinfo{year}{1993}).

\bibitem[{\citenamefont{Boulanger et~al.}(2020)\citenamefont{Boulanger,
  Buisseret, Dehouck, Dierick, and White}}]{PhysRevE.102.062403}
\bibinfo{author}{\bibfnamefont{N.}~\bibnamefont{Boulanger}},
  \bibinfo{author}{\bibfnamefont{F.}~\bibnamefont{Buisseret}},
  \bibinfo{author}{\bibfnamefont{V.}~\bibnamefont{Dehouck}},
  \bibinfo{author}{\bibfnamefont{F.}~\bibnamefont{Dierick}}, \bibnamefont{and}
  \bibinfo{author}{\bibfnamefont{O.}~\bibnamefont{White}},
  \bibinfo{journal}{Phys. Rev. E} \textbf{\bibinfo{volume}{102}},
  \bibinfo{pages}{062403} (\bibinfo{year}{2020}),
  \urlprefix\url{https://link.aps.org/doi/10.1103/PhysRevE.102.062403}.

\bibitem[{\citenamefont{Boulanger et~al.}(2019)\citenamefont{Boulanger,
  Buisseret, Dierick, and White}}]{Boulanger:2018tue}
\bibinfo{author}{\bibfnamefont{N.}~\bibnamefont{Boulanger}},
  \bibinfo{author}{\bibfnamefont{F.}~\bibnamefont{Buisseret}},
  \bibinfo{author}{\bibfnamefont{F.}~\bibnamefont{Dierick}}, \bibnamefont{and}
  \bibinfo{author}{\bibfnamefont{O.}~\bibnamefont{White}},
  \bibinfo{journal}{Eur. Phys. J.} \textbf{\bibinfo{volume}{C79}},
  \bibinfo{pages}{60} (\bibinfo{year}{2019}), \eprint{1811.07733}.

\bibitem[{\citenamefont{White et~al.}(2008)}]{white08}
\bibinfo{author}{\bibfnamefont{O.}~\bibnamefont{White}} \bibnamefont{et~al.},
  \bibinfo{journal}{J Neurophysiol} \textbf{\bibinfo{volume}{100}}
  (\bibinfo{year}{2008}).

\bibitem[{\citenamefont{Nelson}(1983)}]{Nelson83}
\bibinfo{author}{\bibfnamefont{W.}~\bibnamefont{Nelson}},
  \bibinfo{journal}{Biol. Cybern.} \textbf{\bibinfo{volume}{46}},
  \bibinfo{pages}{135} (\bibinfo{year}{1983}).

\bibitem[{\citenamefont{Hogan}(1984)}]{Hogan84}
\bibinfo{author}{\bibfnamefont{N.}~\bibnamefont{Hogan}}, \bibinfo{journal}{J.
  Neurosci.} \textbf{\bibinfo{volume}{4}}, \bibinfo{pages}{2745}
  (\bibinfo{year}{1984}).

\bibitem[{\citenamefont{Hagler}(2015)}]{Hagler2015}
\bibinfo{author}{\bibfnamefont{S.}~\bibnamefont{Hagler}}
  (\bibinfo{year}{2015}), \urlprefix\url{arXiv:1509.06981}.

\bibitem[{\citenamefont{Pais and Uhlenbeck}(1950)}]{Pais:1950za}
\bibinfo{author}{\bibfnamefont{A.}~\bibnamefont{Pais}} \bibnamefont{and}
  \bibinfo{author}{\bibfnamefont{G.~E.} \bibnamefont{Uhlenbeck}},
  \bibinfo{journal}{Phys. Rev.} \textbf{\bibinfo{volume}{79}},
  \bibinfo{pages}{145} (\bibinfo{year}{1950}).

\bibitem[{\citenamefont{Assi et~al.}(2016)\citenamefont{Assi, Bakouny, Karam,
  Massaad, Skalli, and Ghanem}}]{assi}
\bibinfo{author}{\bibfnamefont{A.}~\bibnamefont{Assi}},
  \bibinfo{author}{\bibfnamefont{Z.}~\bibnamefont{Bakouny}},
  \bibinfo{author}{\bibfnamefont{M.}~\bibnamefont{Karam}},
  \bibinfo{author}{\bibfnamefont{A.}~\bibnamefont{Massaad}},
  \bibinfo{author}{\bibfnamefont{W.}~\bibnamefont{Skalli}}, \bibnamefont{and}
  \bibinfo{author}{\bibfnamefont{I.}~\bibnamefont{Ghanem}},
  \bibinfo{journal}{Human movement science} \textbf{\bibinfo{volume}{50}},
  \bibinfo{pages}{10} (\bibinfo{year}{2016}).

\bibitem[{\citenamefont{Peat}(1986)}]{peat}
\bibinfo{author}{\bibfnamefont{M.}~\bibnamefont{Peat}},
  \bibinfo{journal}{Physical therapy} \textbf{\bibinfo{volume}{66}},
  \bibinfo{pages}{1855} (\bibinfo{year}{1986}).

\bibitem[{\citenamefont{Hurov}(1986)}]{hurov}
\bibinfo{author}{\bibfnamefont{J.}~\bibnamefont{Hurov}},
  \bibinfo{journal}{Journal of hand therapy} \textbf{\bibinfo{volume}{22}},
  \bibinfo{pages}{328} (\bibinfo{year}{1986}).

\bibitem[{\citenamefont{Kaparulin et~al.}(2020)\citenamefont{Kaparulin,
  Lyakhovich, and Nosyrev}}]{Kaparulin:2020rqz}
\bibinfo{author}{\bibfnamefont{D.~S.} \bibnamefont{Kaparulin}},
  \bibinfo{author}{\bibfnamefont{S.~L.} \bibnamefont{Lyakhovich}},
  \bibnamefont{and} \bibinfo{author}{\bibfnamefont{O.~D.}
  \bibnamefont{Nosyrev}}, \bibinfo{journal}{Phys. Rev. D}
  \textbf{\bibinfo{volume}{101}}, \bibinfo{pages}{125004}
  (\bibinfo{year}{2020}), \eprint{2003.10860}.

\bibitem[{\citenamefont{Shadmehr and Mussa-Ivaldi}(1994)}]{shad94}
\bibinfo{author}{\bibfnamefont{R.}~\bibnamefont{Shadmehr}} \bibnamefont{and}
  \bibinfo{author}{\bibfnamefont{F.}~\bibnamefont{Mussa-Ivaldi}},
  \bibinfo{journal}{J Neurosci.} \textbf{\bibinfo{volume}{14}},
  \bibinfo{pages}{3208} (\bibinfo{year}{1994}).

\bibitem[{\citenamefont{Augurelle et~al.}(2003)\citenamefont{Augurelle, Penta,
  White, and Thonnard}}]{augu03}
\bibinfo{author}{\bibfnamefont{A.}~\bibnamefont{Augurelle}},
  \bibinfo{author}{\bibfnamefont{M.}~\bibnamefont{Penta}},
  \bibinfo{author}{\bibfnamefont{O.}~\bibnamefont{White}}, \bibnamefont{and}
  \bibinfo{author}{\bibfnamefont{J.}~\bibnamefont{Thonnard}},
  \bibinfo{journal}{Exp Brain Res.} \textbf{\bibinfo{volume}{148}},
  \bibinfo{pages}{533} (\bibinfo{year}{2003}).

\bibitem[{\citenamefont{Cothros et~al.}(2008)\citenamefont{Cothros, Wong, and
  Gribble}}]{cothros}
\bibinfo{author}{\bibfnamefont{N.}~\bibnamefont{Cothros}},
  \bibinfo{author}{\bibfnamefont{J.}~\bibnamefont{Wong}}, \bibnamefont{and}
  \bibinfo{author}{\bibfnamefont{P.~L.} \bibnamefont{Gribble}},
  \bibinfo{journal}{PLOS ONE} \textbf{\bibinfo{volume}{3}}, \bibinfo{pages}{1}
  (\bibinfo{year}{2008}).

\bibitem[{\citenamefont{Bernstein}(1967)}]{Bernstein}
\bibinfo{author}{\bibfnamefont{N.}~\bibnamefont{Bernstein}},
  \emph{\bibinfo{title}{The Coordination and Regulation of Movements}}
  (\bibinfo{publisher}{Oxford: Pergamon Press}, \bibinfo{address}{Oxford},
  \bibinfo{year}{1967}).

\bibitem[{\citenamefont{Fitts}(1954)}]{Fitts}
\bibinfo{author}{\bibfnamefont{P.~M.} \bibnamefont{Fitts}},
  \bibinfo{journal}{Journal of Experimental Psychology}
  \textbf{\bibinfo{volume}{47}}, \bibinfo{pages}{381} (\bibinfo{year}{1954}).

\bibitem[{\citenamefont{Arnold}(1989)}]{Arnold}
\bibinfo{author}{\bibfnamefont{V.~I.} \bibnamefont{Arnold}},
  \emph{\bibinfo{title}{Mathematical methods of classical mechanics}}, Graduate
  Text in Mathematics (\bibinfo{publisher}{Springer-Verlag},
  \bibinfo{year}{1989}), \bibinfo{edition}{second edition} ed.

\end{thebibliography}

\end{document}